\documentclass[aps,pre,twocolumn,showpacs]{revtex4}
\usepackage{epsfig}
\usepackage{times}
\begin{document}

\title{Impact of critical mass on the evolution of cooperation in spatial public goods games}

\author{Attila Szolnoki$^1$ and Matja{\v z} Perc$^2$}
\affiliation
{$^1$Research Institute for Technical Physics and Materials Science,
P.O. Box 49, H-1525 Budapest, Hungary \\
$^2$Department of Physics, Faculty of Natural Sciences and Mathematics, University of \\ Maribor, Koro{\v s}ka cesta 160, SI-2000 Maribor, Slovenia}

\begin{abstract}
We study the evolution of cooperation under the assumption that the collective benefits of group membership can only be harvested if the fraction of cooperators within the group, \textit{i.e.} their critical mass, exceeds a threshold value. Considering structured populations, we show that a moderate fraction of cooperators can prevail even at very low multiplication factors if the critical mass is minimal. For larger multiplication factors, however, the level of cooperation is highest at an intermediate value of the critical mass. The latter is robust to variations of the group size and the interaction network topology. Applying the optimal critical mass threshold, we show that the fraction of cooperators in public goods games is significantly larger than in the traditional linear model, where the produced public good is proportional to the fraction of cooperators within the group.
\end{abstract}

\maketitle

The emergence of cooperation among selfish individuals within the framework of evolutionary game theory is an intensively studied problem \cite{nowak_s06}. While the prisoner's dilemma, snowdrift and the stag-hunt games typically entail pairwise interactions, the public goods game traditionally considers larger groups of interacting players \cite{chen_xj_pre09}. Essentially, however, all mentioned social dilemmas can consider either pairwise or group interactions, as was suggested in Refs.~\cite{bonacich_jcr76, souza_mo_jtb09}. Indeed, it is expected that the possibility of multi-player interactions can bring about phenomena that cannot be observed in case of pairwise interactions, especially when the underlying topology of players is structured rather than well-mixed \cite{szabo_pr07, van-segbroeck_prl09}.

In the classical public goods game setup, individuals engage in multi-player interactions and decide whether they wish to contribute (cooperate) or not (defect) to the common pool. The accumulated contributions, equalling one each, are summoned and multiplied by a factor large than one, \textit{i.e.} the so-called multiplication factor, due to synergy effects of cooperation. Subsequently, the resulting assets are shared equally among all group members, irrespective of their initial contribution to the common pool \cite{brandt_prsb03}. Although the benefits of mutual cooperation, especially if compared to individual or independent cooperative efforts, are widely accepted, they do not apply in all situations. More specifically, the accumulated public good doesn't always depend proportionally on the fraction of cooperators within the group. In the beginning the start-up costs need to be absorbed and offset, therefore decimating the expected return to the initial contributors. On the other hand, when the output limit of a joint venture approaches, the impact of additional contributors becomes marginal \cite{marwell_93}
In extreme situations the sparse occurrence of cooperators in the group makes it impossible to produce public goods. Instead, a minimal number of cooperative contributors is required, \textit{i.e.} the so-called ``critical mass", to elicit the full advantage of group action. There exist several real-life examples supporting such a binary outcome assumption. For example, the building of a bridge (or something that is of value to the majority) within a community requires a certain minimal fraction of supporters. However, if the critical mass of those is not reached, all good aims will go to waste. Group hunting of predators can also be mentioned as an example of a ``gain all-or-nothing" activity. In this work we explore how the size of the critical mass within a group influences the global level of cooperation in a society where the relations between players are defined by spatial interactions \cite{szabo_prl02}.

In the studied public goods game players occupy the nodes of an interaction graph where, for simplicity, every node has the same degree $z$. The focal player forms a group of size $G=z+1$ with its nearest neighbors, although the group size can be extended by considering more distant neighbors as well. Importantly, each player belongs to $G$ different groups, as it is illustrated in Ref.~\cite{santos_n08}. Initially every player on site $x$ is designated either as a defector ($s_x = 0$) or cooperator ($s_x = 1$) with equal probability. The total payoff $P_x$ of player $x$ is the sum of partial payoffs $P_{x,i}$, which are collected from groups around every focal player $i$ where $x$ is also a member ($x \in G_i$). Such a payoff is given by
\begin{equation}
\label{eq:payoff}
P_{x,i}= \cases {\tilde{r} G - s_x, {\rm \, if \,\, TH} \leq \displaystyle \sum_{j \in G_i} s_j \cr \,\,\,\,\,\,\,\,\,\,\,\, - s_x, {\rm \, otherwise} \cr }\;,
\end{equation}
where $\tilde{r}=r/G$ is the normalized multiplication factor originating from the synergy effects of mutual cooperation, and the sum runs over all the players $j$ that are members of the group centered around the focal player $i$. Here $1\leq {\rm TH} \leq G$ denotes the threshold value of the critical mass. More precisely, group members can benefit from the joint venture only if the number of cooperators within a group is equal or exceeds this threshold. In the opposite case the cooperators loose their investments while the defectors gain nothing. A similar assumption was made in earlier works, where the evolution of cooperation in well-mixed populations was studied \cite{bach_jtb06, souza_mo_jtb09}. There a group of players $G$ is chosen randomly, and the mentioned threshold condition is introduced to harvest collective benefits. Due to this a new fixed point emerges where cooperators and defectors can coexist. Souza et al. \cite{souza_mo_jtb09} have shown that the fraction of cooperators in the coexistence regime increases with the critical mass. In our case, however, the possibility of repeated interactions within the realm of structured populations yields a different threshold dependence of the cooperation level, as we will report below.

\begin{figure}
\centerline{\epsfig{file=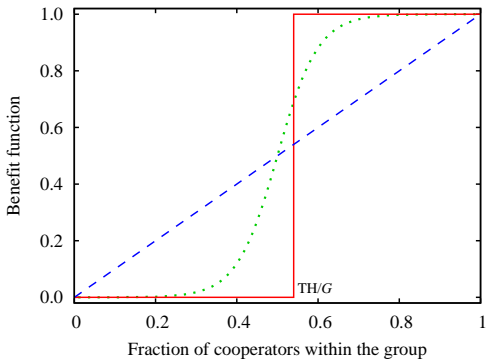,width=6.5cm}}
\caption{(color online) Comparative plots of benefit functions in dependence on the fraction of cooperators within a group. The dotted green ``S"-shaped curve corresponds to the actual profile \cite{marwell_93}, while the linear dependence (dashed blue line) is the one assumed most frequently in public goods games. The step-like gain all-or nothing function (red solid line) is used at present, where group benefits can be harvested only if the critical mass of cooperators exceeds the threshold value (${\rm TH}/G$). For comparisons, all functions are normalized by their maximal values.}
\label{profile}
\end{figure}

To visualize the impact of introducing the critical mass threshold, it is instructive to compare different profiles of actually produced public goods in dependence on the fraction of cooperators within a given group, as shown in Fig.~\ref{profile}. Most commonly, the produced public good is assumed to be directly proportional with the number of cooperators, thus yielding a linear profile (dashed blue line). However, Marwell \textit{et al.} and Heckathorn \cite{marwell_93} argued that such a relation is not necessarily in agreement with actual observations, and that in fact an ``S"-shaped dependence (dotted green line) is much more fitting to reality. The introduction of critical mass yields a simplification, or rather an extreme version of the latter dependence, giving rise to a step-like function (solid red line) going from zero to the maximal value at the threshold (${\rm TH}/G$ in Fig.~\ref{profile}). The saturation beyond the threshold accounts for the fact that the growth of cooperators may not necessarily lead to an enhanced social welfare.

Primarily applied interaction graphs are the square ($z=4$) and the triangle lattice ($z=6$), the two being representative for networks having zero and nonzero clustering coefficient, although our observations were tested on random regular graphs having $z=4$ as well. Different group sizes $G$ are also considered, which we will specify when presenting the results. The applied system size ranged from $10^4 - 10^6$ players. Following the standard dynamics of spatial models, during an elementary Monte Carlo step a player $x$ and one of its neighbors $y$ are selected randomly. After calculating their payoffs $P_x$ and $P_y$ as described above, player $x$ tries to enforce its strategy $s_x$ on player $y$ in accordance with the probability $W(s_x \rightarrow s_y)=1 /\{1+\exp[(P_y-P_x)/K]\}$, where $K>0$ is a noise parameter describing the uncertainty by strategy adoptions \cite{szabo_pr07}. As is natural, better performing strategies are adopted with a large probability, although at nonzero values of $K$ strategies performing  poorly can spread too.
In what follows we will use a fixed value of $K=0.5$ without loss of generality. As it was previously shown, the introduction of multi-player interactions gives rise to a robust topology-independent noise dependence of the cooperation level \cite{szolnoki_pre09c}. During a Monte Carlo step (MCS) all players will have a chance to spread their strategy once on average. The typical relaxation period was up to $2 \cdot 10^4$ MCS before the stationary fraction of cooperators ($f_C$) was evaluated, although substantially faster relaxation times were also observed, as will be described below.

It is important to note that the introduction of critical mass results in a setup that is different from the so-called threshold public goods game \cite{cadsby_pc08}.
In the latter case, players are provided with an endowment and subsequently they must decide how much of that to contribute for the provision of a public good. If the sum of all contributions reaches a threshold, each individual receives a reward. Here, the cooperators contribute a fixed amount, whereafter the constitution of the group determines whether their initial input will be exalted or go to waste. Moreover, threshold public goods games were studied only in well-mixed or single-group populations.

\begin{figure}
\centerline{\epsfig{file=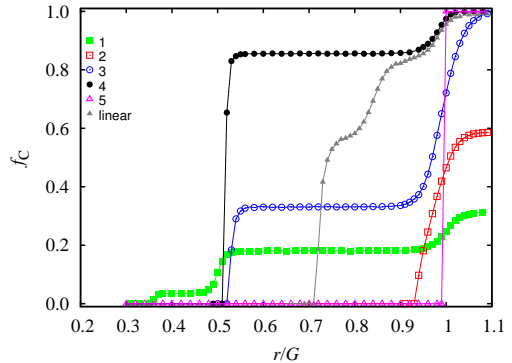,width=6.8cm}}
\caption{(color online) Fraction of cooperators as a function of the normalized multiplication factor $r/G$ for different threshold values. The outcome of the linear model is shown as well. The interaction graph was a square lattice with $G=5$.}
\label{g5}
\end{figure}

Starting with the basic setup entailing the square lattice with $G=5$, we present the fraction of cooperators ($f_C$) as a function of $r$ for different threshold values in Fig.~\ref{g5}. First, it can be observed that using a minimal critical mass for the threshold (${\rm TH}=1$), it is possible to sustain a small fraction of cooperators even if the multiplication factor is extremely low [for comparison, note that defectors always dominate completely below $r/G=0.7$ (see Fig.~3 in Ref.~\cite{szolnoki_pre09c}) when the linear model is used]. At such low $r$ values, the modest total amount of produced public goods is supplied by a single cooperator within every group. Consequently, the frequency of cooperators is proportional to $G^{-2}$ (one cooperator per $G$-sized group, whereby every cooperator is a member of $G$ groups). Secondly however, when $r$ is increased, the advantage of aggregated cooperators can be utilized more efficiently only at larger threshold values (${\rm TH}>1$). Yet the increase in the overall cooperation level for intermediate values of $r$ cannot be sustained if the critical mass becomes too high, thus suggesting the existence of an optimal threshold for the evolution of cooperation.

\begin{figure}
\centerline{\epsfig{file=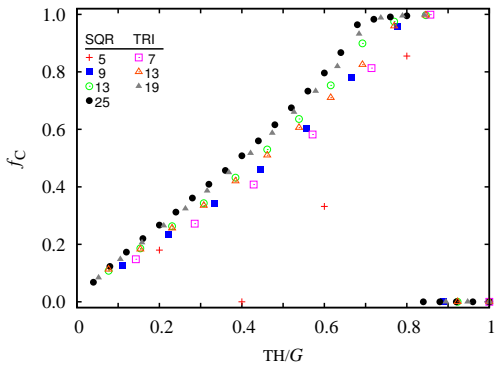,width=6.6cm}}
\caption{(color online) Fraction of cooperators as a function of the normalized threshold value ${\rm TH}/G$ for different group sizes and interaction graphs (SQR=square lattice; TRI=triangular lattice) at $r/G=0.6$. Note that the normalization of ${\rm TH}$ and $r$ with $G$ is essential for relevant comparisons.}
\label{collapse}
\end{figure}

To explore the robustness of our observations we have also used larger group sizes $G$, thereby gaining the advantageous possibility of fine-tuning the threshold value more precisely. Specifically, the applied group sizes were $G=9$, $13$ and $25$ for the square lattice, and $G=7$, $13$ and $19$ for the triangular lattice. Figure~\ref{collapse} shows the results, indicating clearly the existence of an optimal intermediate critical mass for which the cooperation level is highest, independently from the group size or the underlying interaction graph.

The robust existence of an optimal critical mass can be explained if we distinguish cooperators based on whether their initial contributions are exalted, hence increasing the produced public good, or go to waste. Depending on this, we designate cooperators accordingly as being either ``active" or ``inactive". An inactive cooperator is always vulnerable in the presence of defectors because the moderate aggregation of other cooperators in its vicinity is insufficient for spatial reciprocity to work \cite{nowak_s06}. This happens frequently if the threshold is set too high, having as the inevitable consequence the fast extinction of the cooperative strategy. In the opposite limit, \textit{i.e.} when the threshold is very low, practically all cooperators are  active. Then, however, the cooperators don't have a strong incentive to aggregate because an increase in their density will not notably elevate their individual fitness. Consequently, in this case only a moderate fraction of cooperators coexists with the prevailing defectors. At intermediate thresholds the status of cooperators may vary depending on their location on the graph. In particular, there are places where their local density exceeds the threshold, and thus the cooperators there are active. These cooperators can prevail efficiently against defectors. Yet there are also places where the cooperators are inactive because their density is locally insufficient. In these areas defectors can easily defeat cooperators. Importantly, however, after the initial reconfigurations the emerging domains of active cooperators start spreading prolifically in the sea of defectors and are ultimately victorious. The final cooperation level obviously depends also on the multiplication factor, whereby this dependence is similar as was reported in previous works employing the linear public goods function \cite{szabo_prl02, szolnoki_pre09c}.

\begin{figure}
\centerline{\epsfig{file=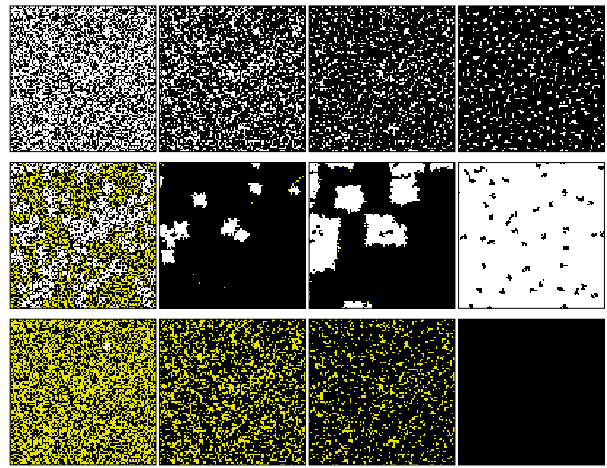,width=8cm}}
\caption{(color online) Time evolution (from left to right) of an identical random initial state on a square lattice having $G=25$ for ${\rm TH}=2$ (top row), $17$ (middle row) and $22$ (bottom row), at $r/G=0.6$. Black are defectors, while white and yellow (light grey) are active and inactive cooperators, respectively. Note that the partly different coloring in the first column is due to the differences in status of some cooperators appearing as a consequence of different ${\rm TH}$ values. All panels show a $100 \times 100$ excerpt of a larger $400 \times 400$ lattice.}
\label{snapshots}
\end{figure}

Figure~\ref{snapshots} demonstrates the preceding argumentation effectively. It shows how the system evolves for three different representative threshold values on a square lattice with $G=25$. The thresholds are ${\rm TH}=2$ (top row), $17$ (middle row) and $22$ (bottom row). Time evolution goes from the left towards the right snapshots, starting with the random initial state and ending with the stationary state. Black color is used for defectors, while white and yellow depict active and inactive cooperators, respectively. It is interesting to note that, despite appearances, the leftmost panels depicting the initial state are completely identical (exactly the same random initial conditions were used). Importantly, however, the application of different threshold values yields an adverse classification of cooperators on those that are active (white) and those that are inactive (yellow), which obviously has an impressive impact on the final state (compare the rightmost snapshots).

When the threshold is low (top row of Fig.~\ref{snapshots}) practically all cooperators are active, thus supplying their groups with the maximal payoff. As we have argued above, in this case a higher density of cooperators would not be advantageous. Therefore the active cooperators (colored white) don't aggregate. Of course, the stationary fraction of cooperators depends on the actual value of ${\rm TH}$, whereby interestingly the resulting cooperation level is larger than the applied threshold value. The difference between $f_C$ and ${\rm TH}/G$ becomes relevant when ${\rm TH}$ approaches the optimal value. If the imposed critical mass is too high (bottom row of Fig.~\ref{snapshots}), the vast majority of cooperators becomes inactive (colored yellow). Despite the fact that the interactions amongst players are structured, \textit{i.e.} the underlying graph is a lattice, the spatial reciprocity cannot work and thus the cooperators go extinct very fast (around $10^2$ MCSs suffice to get an absorbing $D$ phase even for large system sizes). The leftmost bottom snapshot shows clearly that only a tiny fraction of nearby cooperators can initially exceed the necessary threshold (small white area). However, they cannot propagate because their spreading would require too many defectors changing their strategy in the vicinity of the border. Oppositely, the strategy change of a single active cooperator can easily decrease their density below the critical mass threshold, which leads the defectors to full dominance, as shown in the rightmost bottom snapshot.

In the intermediate threshold region (middle row of Fig.~\ref{snapshots}), we can observe a fast extinction of inactive cooperators. Because of the moderate critical mass, however, a new phenomenon emerges. Active cooperators (colored white) can easily protect themselves against the invasion, and more importantly still, they can also alter their neighborhoods and therefore spread in the sea of defectors. Eventually this process results in a highly cooperative stationary state, as shown in the rightmost middle snapshot. In fact, the introduction of an intermediate critical mass paves the way for spatial reciprocity to work extremely effectively, leading to the selection of the most beneficial state in the course of the evolution. If compared to the extinction of inactive cooperators the mentioned process is slower because it relies on a propagation mechanism. From the defector's point of view, however, the negative feedback effect due to their own spreading is more severe than in the traditional linear public goods game (see dashed blue line in Fig.~\ref{profile}). In particular, while in the presently proposed critical mass model the invasion of defectors may result in a sudden loss of collective benefits, the linear model always ensures a small amount of public goods in the vicinity of cooperators. This is why the fraction of defectors remains at a very low level, even for small multiplication factors, if the optimal critical mass threshold is imposed.

In sum, we have shown that the evolution of cooperation in spatial public goods games can be promoted effectively, even at unfavorable conditions (\textit{i.e.} low $r$ values), via the introduction of critical mass acting as a threshold for initial contributions to the common pool. In contrast with well-mixed populations, here the impact of critical mass is optimal at an intermediate value of the threshold, which allows spatial reciprocity to work more effectively than in the linear public goods game. Notably, the optimal critical mass was found to be robust against variations in the group size and the underlying interaction network. The revealed mechanism for the promotion of cooperation can be understood by taking into account the binary (active/inactive) impact of cooperators, which emerges spontaneously depending on their local density. In future studies, it will be interesting to investigate how locally diverse values of critical mass influence the global level of cooperation, and more generally, if and how a coevolutionary model \cite{perc_bs10}, where besides the strategy adoptions of players groups will also be able to adopt the critical threshold value from a more successful community, can be devised so that the optimal thresholds are selected naturally.

The authors acknowledge support from the Hungarian National Research Fund (grant K-73449) the Bolyai Research Grant, the Slovenian Research Agency (grant Z1-2032), and the Slovene-Hungarian bilateral incentive (grant BI-HU/09-10-001).

\end{document}